\documentclass{elsart3p}

\usepackage{natbib}

 \usepackage{graphics}
 \usepackage{graphicx}
 \usepackage{epsfig}

\usepackage{amssymb}

\begin{document}

\begin{frontmatter}




\title{Evidence of longterm cyclic evolution of radio pulsar periods\thanksref{label1}}
\thanks[label1]{This work has been supported by the Russian Foundation for Basic Research 
(grant No 04-02-17555), Russian Academy of Sciences (program "Evolution 
of Stars and Galaxies"), and by the Russian Science Support Foundation. The
authors are also grateful to the anonymous reviewers for their valuable
comments.}

\author[adr1]{Anton Biryukov}, 
\ead{eman@sai.msu.ru}
\author[adr2]{Gregory Beskin},
\ead{beskin@sao.ru} 
\author[adr2]{Sergey Karpov},
\author[adr1]{Lisa Chmyreva} 
\address[adr2]{Special Astrophysical Observatory of RAS, Nizhniy
Arkhyz, Karachaevo-Cherkessia, Russia, 369167}
\address[adr1]{Sternberg Astronomical Institute of MSU, 13, Universitetsky
pr.,Moscow, Russia, 119992}

\begin{abstract}
The measurements of pulsar frequency second derivatives have shown that they 
are $10^2-10^6$ times larger than expected for standard pulsar spin-down law. 
Moreover, the second derivatives as well as braking indices are even negative
for about half the pulsars. We explain these paradoxical results on the basis 
of the statistical analysis of the rotational parameters $\nu$, $\dot \nu$ 
and $\ddot \nu$ of the subset of 295 pulsars taken mostly from the ATNF 
database. We have found a strong correlation between $\ddot \nu$ and 
$\dot \nu$  for both $\ddot\nu > 0$ (correlation coefficient $r\approx0.9$) 
and $\ddot\nu < 0$ ($r\approx0.85$), as well as between $\nu$ and $\dot\nu$ 
($r\approx0.6\div0.7$). We interpret these dependencies as evolutionary ones due
to 
$\dot\nu$ being nearly proportional to the pulsars' age. 

The derived statistical relations as well as ``anomalous'' values of 
$\ddot\nu$ are well described by assuming the existence of long-time variations of the spin-down rate. The pulsar frequency evolution, therefore, consists of 
secular change of $\nu_{ev}(t)$, $\dot\nu_{ev}(t)$ and $\ddot\nu_{ev}(t)$ 
according to the power law with $n\approx5$, the irregularities, observed 
within the timespan as timing noise, and the non-monotonous variations on the 
timescale of several tens of years, which is larger than that of the timespan. It is possible 
that the nature of long-term variations is similar to that of short-term ones.
The idea of non-constant secular pulsars' braking index $n$ is also analysed.
\end{abstract}


\end{frontmatter}

\section{Introduction}
\label{intro}
The spin-down of radio pulsars is caused by the conversion of their rotation 
energy into emission. According to the ``classical'' approach, their rotational 
frequencies $\nu$ evolve obeying the spin-down law $\dot\nu=-K\nu^n$, where 
$K$ is a positive constant that depends on the magnetic dipole moment and the 
moment of inertia of the neutron star, and $n$ is the braking index. The 
latter can be determined observationally from measurements of $\nu$, 
$\dot \nu$ and $\ddot \nu$ as $n={\nu \ddot \nu}/{\dot \nu^2}$. For a simple 
vacuum dipole model of the pulsar magnetosphere $n=3$; the pulsar wind decreases 
this value to $n=1$; for multipole magnetic field $n \ge 5$ \citep{man77}. At 
the same time the measurements of pulsar frequency second derivatives 
$\ddot \nu$ have shown that their values are much larger than expected for 
standard spin-down law and are even negative for about half of all pulsars. 
The corresponding braking indices range from $-10^6$ to $10^6$ 
\citep{d'a93, chu03, hob04}.

It was found that the significant correlations between $|\ddot\nu|$ 
($|\ddot P|$) and $\dot\nu$ ($\dot P$) demonstrate the fact that 
the absolute values of the $\ddot \nu$ and $\ddot P$ are
larger for younger (with large $|\dot \nu|$) pulsars \citep{cor85, arz94, lyn99}.
The anomalously high 
and negative values of $\ddot\nu$ and $n$ may be interpreted as a result of 
low-frequency components of the ``timing noise'' -- a complex change of 
pulsars' rotational phase within a timespan \citep{d'a93}. Or, as a result
of any long-term influence on the pulsars' spin-down. \citep{gul77, dem79}

It is clear that the timespan of observations is by no means intrinsic to
pulsar physics.  Indeed, the variations of rotational parameters may take 
place on larger timescales as well. However, the timescale of observations 
naturally divides these variations into two separate classes of
manifestations: (i) the well-known ``timing noise'' -- the residuals in respect
to the best fit for the timing solution and (ii) the ``long term timing noise''
-- the systematic shift of the best fit coefficients (i.e. in the measured
values of $\nu$, $\dot\nu$, $\ddot\nu$) relative to some mean or 
expected value from the model.

Up to date we know nearly 200 pulsars for which the timespan of observations 
is greater than 20 years, and the values of their $\ddot \nu$ still turn out 
to be anomalously large \citep{hob04}.

For example, for the {\it PSR B1706-16} pulsar, variations of $\ddot \nu$ 
with an amplitude of $10^{-24}$ s$^{-3}$ have been detected on a timescale of
several years (see Fig.7 in \cite{hob04}), with the value of $\ddot\nu$ 
depending on the time interval selected. However, the fit over the entire 25 
year timespan gives a value of $\ddot\nu=3.8\cdot10^{-25}$ s$^{-3}$ with a 
few percent accuracy (which leads to a braking index $\approx 2.7\cdot10^3$). 

In the current work we provide observational evidence of the non-monotonous
evolution of pulsars on timescales larger than the typical contemporary 
timespan of observations (tens of years), using the statistical analysis of
the measured $\nu$, $\dot\nu$ and $\ddot \nu$. We estimate the main parameters
of such long-term variations and discuss their possible relation to the 
low-frequency terms of timing noise. We have also derived the parameters 
of pulsar secular spin-down. We plotted the $\nu-\dot\nu$ diagrams for
295 and for 1337 ``ordinary'' radiopulsars. The smaller subset consists of pulsars with
measured $\ddot \nu$. The bigger one -- of all the ``ordinary'' pulsars with
measured $\nu$ and $\dot\nu$, taken from the ATNF
database \citep{man05}. We have found a good
correlation beetween these two parameters and determnined the mean slope of
the $\dot\nu-\nu$ distribution, which is in agreement with $n \sim 5$.

\section{Statistical analysis of the ensemble of pulsars}
\label{statan}
As was stated above, earlier works have shown 
the possibility of long-term variations of pulsars' rotational
frequency. Our statistical analysis shows that the measurements of most the $\ddot \nu$
reflect the pulsars' spin-down evolution on a 
timescale larger than the duration of observations, and uses the 
parameters of 295 pulsars.
\begin{figure}[t]
{\centering \resizebox*{1\columnwidth}{!}{\includegraphics[angle=270]
{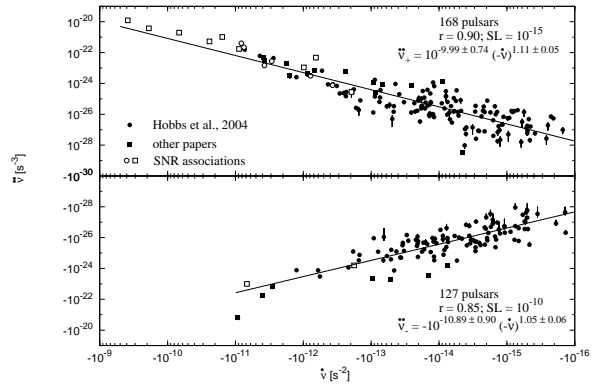}} \par}
\caption{The $\ddot \nu - \dot \nu$ diagram for 295 pulsars. The figure 
shows the pulsars taken from \citep{hob04} as circles, and the objects 
measured by other groups as squares. Open symbols represent the relatively 
young pulsars associated with supernova remnants.
Analytical fits for both positive and negative branches are shown as solid 
lines. Measurment errors are shown as error bars.}
\label{fig_diagram}
\end{figure}
From the 389 objects of the ATNF catalogue \citep{man05} with known $\ddot \nu$
we compiled a list of ``ordinary'' radio pulsars with $P>20$ ms, 
$\dot P >10^{-17}$ s/s, and with relative accuracy of second derivative
measurements better than 75\%. We excluded recycled, anomalous and binary 
pulsars. 26 supplementary pulsars from other sources 
\citep{d'a93, chu03} were added. The parameters of all pulsars were plotted on the 
$\ddot \nu - \dot \nu$ diagram (Fig. \ref{fig_diagram}).

The basic result of the statistical analysis of this data is a significant
correlation between $\ddot \nu$ and $\dot \nu$, both for 168 objects with $\ddot
\nu > 0$ (correlation coefficient $r\approx0.90$) and for 127 objects with
$\ddot \nu < 0$ ($r\approx0.85$). Both groups follow nearly linear laws,
however they are not exactly symmetric relative to $\ddot \nu=0$. We divided
both branches into 6 intervals of $\dot \nu$, computed the mean values and their
standard deviations of $\ddot \nu_{\pm}$ for each interval. We rejected the hypothesis
of the branches symmetry with a 0.04 significance level. Also, the absolute
values of $\ddot \nu_{+}$ are systematically larger than the corresponding $\ddot \nu_{-}$
(the difference is positive in 5 intervals out of 6) and the difference of analytical
fits to branches is positive over the $-10^{-11}\div-10^{-15}$ $s^{-2}$ interval of
$\dot \nu$. These are the arguments in favour of a small positive assymetry of
the branches.

\begin{figure}[t]
{\centering \resizebox*{1\columnwidth}{!}{\includegraphics[angle=270]
{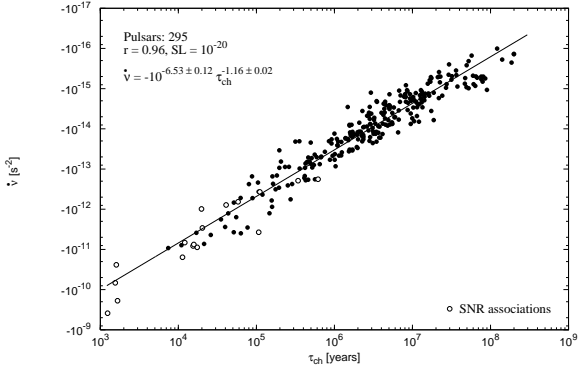}} \par}
\caption{The obvious $\dot \nu - \tau_{ch}$ dependence for pulsars with 
the measured second derivative. The open symbols represent pulsars 
associated with SNRs. The solid line represents the best fit, the dotted 
lines -- 1-$\sigma$ range. The index of power $-1.16\pm0.02$ is 
significantly different from $-1.0$ which indicates the presence of the 
$\dot\nu - \nu$ correlation.}
\label{fig_taudnu}
\end{figure}

We found an obvious correlation of $\dot \nu$ with the characteristic age
$\tau_{ch} = -\frac12 \frac{\nu}{\dot \nu}$ ($r=0.96$, see Fig. \ref{fig_taudnu}):
\begin{equation}
\dot\nu=-10^{-6.53\pm0.12}\tau_{ch}^{-1.16\pm0.02}
\label{eq_tau_dnu}
\end{equation}
 This correlation will be discussed below in the light of the $\dot \nu - \nu$
dependency. The main feature of (\ref{eq_tau_dnu}) is that the index of power
$-1.16 \pm 0.02$ is significantly different from $-1.0$. It is obvious that
$\dot\nu$ would be $\sim \tau_{ch}^{-1}$ if $\nu$ and $\dot \nu$ were completely uncorrelated.
Therefore, the derived result provides an argument in favour of the $\dot
\nu-\nu$ correlation.

\begin{figure}[t]
{\centering \resizebox*{1\columnwidth}{!}{\includegraphics[angle=270]
{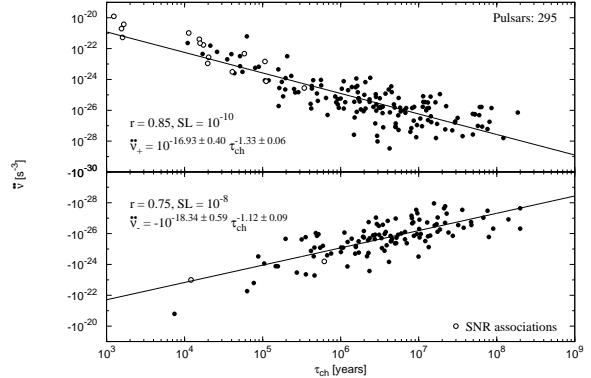}} \par}
\caption{The $\ddot \nu - \tau_{ch}$ dependence for pulsars with 
the measured second derivative. The open symbols represent pulsars 
associated with SNRs. This diagram is an argument in favour of the idea
that pulsars evolve from large to small $|\ddot \nu|$ values
independently from the $\ddot \nu$ sign.}
\label{fig_ddotnutau}
\end{figure}

The $\dot \nu$ and $\tau_{ch}$ are nearly 
proportional, which leads to a significant correlation of $\tau_{ch}$ both 
with $\ddot\nu$ ($r=0.85$ for the positive branch and $r=0.75$ for the 
negative one, Fig. \ref{fig_ddotnutau}) and with $n$ ($r=0.75$ and $r=0.76$
correspondingly, Fig. \ref{fig_ntau}).

The correlations found are fully consistent with the results published in 
\citep{cor85, arz94, lyn99}, as well as in \citep{ura06}. However, the 
branches with $\ddot \nu >0$ and $\ddot \nu < 0$ in those works were not 
analysed separately from each other (not as $|\ddot \nu|$).

Young pulsars confidently associated with supernova remnants are 
systematically shifted to the left in  Fig. \ref{fig_diagram} (open symbols). The order of magnitude of their physical ages roughly corresponds to that of their 
characteristic ages. This means that any dependence on $\dot \nu$ or 
$\tau_{ch}$ reflects the dependence on pulsar age.
The $\ddot\nu - \dot \nu$ diagram (Fig. \ref{fig_diagram}) may be 
interpreted as an evolutionary one. In other words, each pulsar during its 
evolution moves along the branches of this diagram while increasing the value 
of its $\dot \nu$ (which corresponds to the increase of its characteristic 
age). However, there is an obvious contradiction: on the negative branch, 
$\dot \nu$, being negative, may only decrease with time (since $\ddot \nu$ is 
formally the derivative of $\dot \nu$), and the motion along the negative 
branch may only be backward! This contradiction is easily solved by assuming 
non-monotonous behaviour of $\ddot\nu(t)$, which has an irregular component 
($\delta \ddot \nu$) along with the monotonous one ($\ddot \nu_{ev}$), where 
the subscript ``ev'' marks the evolutionary value.

The characteristic timescale $T$ of such variations must be much shorter than 
the pulsar life time and at the same time much larger than the timescale of 
the observations. As it evolves, a pulsar repeatedly changes sign of 
$\ddot \nu$, in a spiral-like motion from branch to branch, and  spends 
roughly half its lifetime on each one. The asymmetry of the branches reflects 
the positive sign of $\ddot \nu_{ev}(t)$, and therefore, secular increase of 
$\dot\nu_{ev}(t)$ (i.e. all pulsars in their secular evolution move to the 
right on the $\ddot\nu - \dot \nu$ diagram). The systematic decreasing of the branches separation reflects the decreasing of the variations amplitude and/or 
the increasing of its characteristic timescale. 

Any well known non-monotonous variations of $\dot\nu(t)$, like glitches, 
microglitches, timing noise or precession, will manifest themselves in a 
similar way on the $\ddot\nu - \dot \nu$ diagram and lead to extremely high 
values of $\ddot \nu$ \citep{she96, sta00}. However, their characteristic
timescales vary from weeks to years, and they are detected immediately. But 
here the variations on much larger timescales are discussed, and their study 
is possible only statistically, assuming the ergodic behaviour of the ensemble
of pulsars.

In addition, the apparent branches separation on the $\ddot\nu - \dot\nu$ diagram is only due to the logarithmic scale of the plot. The spread
of $\ddot \nu$ values in each branch reaches 3 orders of magnitude, i.e. the
pulsars on the diagram cover almost fully the range of possible $\ddot \nu$
values (for each value of $\dot \nu$). Moreover, we observe the lack
of pulsars near $\ddot \nu \sim 0$
because (i) the present accuracy of the $\ddot \nu$ measurements is no better than 
$10^{-29}$ $s^{-3}$ and (ii) in this area pulsars move faster than anywhere,
because here $\dot \nu \approx const$, $\ddot \nu$ changes its sign and hence
the third derivative of $\nu$ has an extremum.

\section{Non-monotonous variations of pulsar spin-down rate on large timescales}
\label{nonmon}
Variations of the pulsar rotational frequency may be complicated -- periodic, 
quasi-periodic, or completely stochastic. Generally, it may be described as a 
superposition 
\begin{equation}
\nu(t)=\nu_{ev}(t) + \delta\nu(t),
\label{eq1}
\end{equation}
where $\nu_{ev}(t)$ describes the secular evolution of pulsar parameters and 
$\delta\nu(t)$ corresponds to the irregular variations. Similar expressions 
describe the evolution of $\dot \nu$ and $\ddot \nu$ after a differentiation.
$\delta\ddot\nu(t)$ satisfies the obvious condition of zero mean value:
\begin{equation}
<\delta\ddot\nu(t)>_{t} = 0
\end{equation} 
over the timespans larger than the characteristic
timescale of the variations. The amplitude of the observed variations of 
$\ddot\nu$ is related to the dispersion of this process as 
\begin{equation}
\sigma_{\delta\ddot\nu} = A_{\ddot\nu} = \sqrt{<(\delta\ddot\nu)^2>}
\end{equation}

The second derivative values on the upper $\ddot\nu_{+}$ and lower 
$\ddot\nu_{-}$ branches in Fig. \ref{fig_diagram} may be approximately described as
\begin{equation}
\ddot\nu_{\pm}(t) = \ddot\nu_{ev}(t) \pm A_{\ddot\nu}(t) 
\end{equation}
for each pulsar. 
This equation describes an "average" pulsar, while the spread of points 
inside the branches reflects the variations of individual parameters over the 
pulsar ensemble and reaches 4 orders of magnitude.

The second derivative $\ddot \nu$ is the only parameter significantly 
influenced by the timing variations (see Section \ref{disc}). Thus one can 
assume that the measured values of $\nu$ and $\dot \nu$ may be considered to 
be evolutionary, $\nu_{ev}$ and $\dot \nu_{ev}$ (since $\delta \nu$ and 
$\delta \dot \nu$ are small).

\subsection{$\dot\nu - \nu$ diagram}

\begin{figure}[t]
{\centering \resizebox*{1\columnwidth}{!}{\includegraphics[angle=270]
{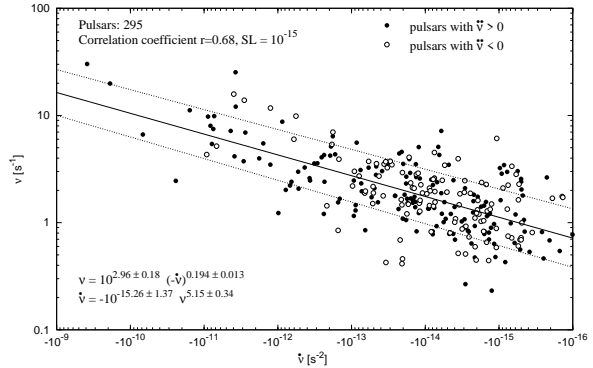}} \par}
\caption{The $\dot \nu - \nu$ diagram for the pulsars with the measured
second derivative. The filled symbols are objects with positive $\ddot\nu$,
the open ones -- with negative $\ddot\nu$. The behaviour of both subsets is the same.
The solid line represents the best fit corresponding to the braking index of $n\approx5$, the dotted lines represent the 1-$\sigma$ range.}
\label{fig_dotnunu}
\end{figure}
Using the relations described above, the secular behaviour $\nu(t)$ (or, 
$\nu(\dot\nu)$) may be found by plotting the studied pulsar group onto the 
$\dot \nu - \nu$ diagram (Fig. \ref{fig_dotnunu}). The objects with $\ddot \nu > 0$ 
and $\ddot \nu < 0$ are marked as filled and open circles, correspondingly. 
It is easily seen that the behaviour of these two sub-groups is the same, 
which is in agreement with the smallness of the pulsar frequency variations 
in respect to the intrinsic scatter of $\nu(\dot \nu)$. However, a strong 
correlation between $\nu$ and $\dot\nu$ ($r\approx0.7)$ is seen, and
\begin{equation}
\dot\nu=-C\nu^n,
\label{eq2}
\end{equation}
where $C=10^{-15.26\pm1.38}$ and $n=5.15\pm0.34$. So, the secular evolution 
of an ``average'' pulsar goes according to the ``standard'' spin-down law with 
$n\approx5$! This result is very interesting on its own, especially since 
the $\nu$ and $\dot \nu$ are always measured as independent values. 

Note that the width of the fit on Fig. \ref{fig_dotnunu} is quite small -- 
only about 1.5 orders of magnitude. If the spin-down is even approximately 
close to being described by the vacuum dipole model, then $C$ should scale as 
$(B_{0} \sin \chi)^2$, where $B_{0}$ is the polar field and $\chi$ is the 
magnetic inclination angle. But the range of $(B_{0}\sin\chi)^2$ over the 
pulsar population is expected to be of many orders of magnitude. Therefore, 
Fig. \ref{fig_dotnunu} once again shows that a simple vacuum dipole model is not 
adequate to observations. 

It is important that there is no contradiction between the derived value
of $n \approx 5$ and values
of the braking index derived in the usual manner using $\ddot \nu$ (these values
lie in the range from approximately $-10^6$ to $10^6$). The measured $\nu$ and
$\dot \nu$ are weakly affected by the timing noise, so
the pulsars move along the $\dot\nu - \nu$ distribution mostly according to the
secular
component of $\ddot \nu$. At the same time the combination 
$\ddot \nu \nu / \dot \nu^2 = n$ includes both the secular and irregular
components of $\ddot \nu$, which leads to anomalous braking indices. 

\begin{figure}[t]
{\centering \resizebox*{1\columnwidth}{!}{\includegraphics[angle=270]
{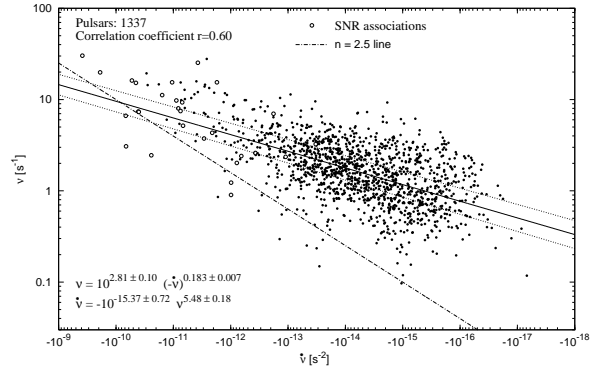}} \par}
\caption{The $\dot \nu - \nu$ diagram for the 1337 pulsars from the ATNF
database.
The distribution shows quite a strong correlation (r = 0.60). The slope of
the distribution corresponds to $n = 5.48 \pm 0.18$. The slope of the
dash-dot line on the plot corresponds to $n = 2.5$. It is clear that if the
values of $n$ for young pulsars are close to $2.5 \div 3.0$, then they should
change to higher values as the pulsars evolve.}
\label{fig_f1f0full}
\end{figure}

The braking index $n \approx 5$ allows us to describe the pulsars
spin-down as ``quadrupole-like''. But it is unlikely that such spin-down is
due to the simple quadrupole structure of the NS magnetic field and the
corresponding radiation. Indeed, in case of vacuum approximation, for the
rotating quadrupole with magnetic field strength
$B_0(R_{NS}) \sim 10^{12}$ $G$ (where $R_{NS} \sim 10^6$ $cm$),
the value of the parameter $C$ from Eq. (\ref{eq2}) would be
as small as $10^{-26}$ \citep{man77, kro91}, which is ten orders of
magnitude smaller than the values measured for the 295 and 1337 pulsars
(see Figures \ref{fig_dotnunu} and \ref{fig_f1f0full}). Moreover, the
total contribution of all the high order multipole components (quadrupole etc.)
is much smaller than that of the dipole component. 

Howewer, it is clear that the described simple quadrupole spin-down
does not satisfy the modern concept of the pulsars' magnetospheres not being vacuum \citep{gj69, man77, bes93}.

A number of papers were published in the last years, where the possibility of
quadrupole components of pulsars' magnetic fields was discussed
(e.g. \cite{zane06}). Thus, at the same time, there is little reason to fully
reject the hypothesis of a more complex pulsars' magnetic fields structure than 
only dipolar.

An additional argument in favour of the $\dot\nu-\nu$ correlation is,
as has been stated above, the dependence between $\dot \nu$ and $\tau_{ch}$,
where the index of power is not equal to $-1.0$. Indeed, since
$\tau_{ch} = -\frac12 \frac{\nu}{\dot\nu}$, and if the $\dot\nu-\nu$ correlation
is absent, then $\nu$ may only bring in some scatter to the $\dot\nu-\tau_{ch}$
dependence without its slope changing (on a logarithmic scale). But, if
$\dot \nu$ and $\nu$ are really correlated, and the slope (braking index) is
about 5, the $\dot\nu-\tau_{ch}$ dependence should show a slope of about $-1.2$
which is roughly consistent with the measured value (see Eq. \ref{eq_tau_dnu}).
Precisely, the slope of the $\dot\nu - \tau$ fit, $\alpha$, and $n$ are related
by the equation: 
\begin{equation}
n = \frac{\alpha}{\alpha - 1}
\end{equation}
So it clearly seen that $n$ is strongly dependent on $\alpha$. The value of
$n$ derived from $\alpha = 1.16 \pm 0.02$ is $7.25 \pm 0.78$. It differs from
$5.15 \pm 0.34$ on a $2.5\sigma$ significance level. This value is less than the
standard $3\sigma$ but close to it.

In a framework of pulsar spin-down analysis we ploted the $\dot\nu -\nu$
diagram
for the 1337 pulsars (Fig. \ref{fig_f1f0full}) taken from the ATNF pulsar
database \citep{man05}. This
subset
does not include recycled, binary and anomalous pulsars, and satisfies the criteria 
described in section (\ref{statan}). The braking index derived from the plot
is $n = 5.48 \pm 0.18$. This value is in a good agreement with the result
described above. The precision of the slope
measurement on Fig. \ref{fig_f1f0full} is so high mostly due
to the large number of pulsars used.

The mean slope of the fit for the $\dot\nu - \nu$ distribution represents
the average value of $n$ for pulsars during their lifetime ($\approx 5$). Up to
date there are several young pulsars with measured braking indices that are both
accurate and precise. These are all smaller than but close to 3. The
fact that the measured value of $n$ was found to be greater than 3, could mean
that pulsars do not evolve with a constant $n$ $(\le 3)$.

Indeed, if the pulsars are really born with $n \le 3$, and assuming they do not
change this value during their evolution, then as they grow older, they should
appear below the main distribution on the
$\dot\nu -\nu$ diagram: approximately in the area
with $\dot \nu \sim -10^{-16} \div -10^{-15}$ $s^{-2}$ and $\nu \sim 0.1$ $Hz$.

As an example, the line with a slope which corresponds to $n = 2.5$ is shown
on Fig. \ref{fig_f1f0full}.  This line significantly deviates from the
main distribution in the area of the oldest pulsars.
If the pulsars were to evolve along this line, they
would arrive in the area below the main distribution as they aged.

Thus, the pulsars start their evolution with $n \le 3$, which seems to be
typical, and are likely to change
their $n$ to higher values as they evolve. Therefore, the
authors consider to be reasonable the idea of
young and old pulsars evolving with different values of $n$. These values are smaller for younger pulsars and larger for older ones. 

Such changing of $n$ can be explained, for example, by the changing of the
dominating spin-down mechanism. In general, a number of
scenarios were proposed, where pulsars' braking index would increase with time
(see for example \cite{cor98} and references therein; see also \cite{rud06}).

On the other hand, the increasing of the braking index may be explained
by the evolution of the parameter $C$ from Eq. (\ref{eq2}). In
case of a simple dipole spin-down $C \propto (B_0 \sin\chi)^2$ and it is not
constant for pulsars during their lifetimes. The magnetic field
decay and magnetic inclination angle evolution ($\dot \chi < 0$, see
\cite{dav70}) will lead to the decreasing of $C$ for the old pulsars. Hence,
the absolute values of $\dot \nu$ will be systematically smaller
than the ``unshifted'' ones (when $C = const$). Thus the slope of the $\nu-\dot
\nu$ distribution (braking index) will increase.

A similar situation also takes place in the model of electric current
spin-down \citep{bes93}. In this case, the relation $\dot \nu = -K\nu^3$ is also true, but
$K \propto (B_0\cos\chi)^2$ with $\dot \chi > 0$, instead of 
$(B_0\sin\chi)^2$ with $\dot \chi < 0$.

In general, any pulsars' spin-down law should depend
on the polar magnetic field $B_0$ and, very likely, on the
magnetic inclination angle $\chi$. Therefore, the idea described
above should remain valid. If so, then the measured $n\approx 5$ means
that pulsars' intrinsic braking index may be {\it less} than 5.

The provided analysis strongly suggests the
presence of a $\dot\nu - \nu$ correlation for ordinary pulsars with a mean
slope (braking index) close to 5. Younger pulsars seem to evolve with a
lower $n$ than the older ones. The pulsars' magnetic field may have
a more complex structure than dipolar, however there is little reason to
insist that $n \approx 5$ is due to a simple quadrupole spin-down. 
It may be a result of the pulsars' magnetic
field decay and/or the evolution of the magnetic inclination angle.

\subsection{Some numerical results}
\label{sub_simmod}
\begin{figure}[t]
{\centering \resizebox*{1\columnwidth}{!}{\includegraphics[angle=270]
{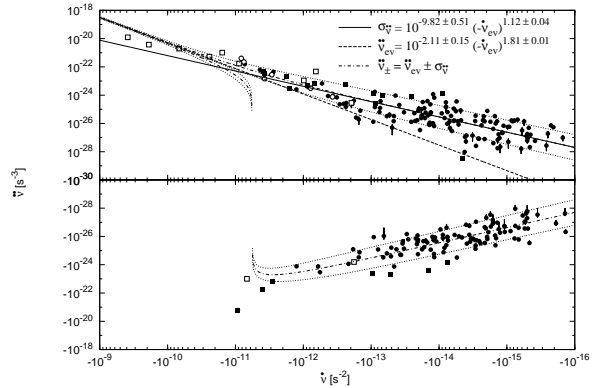}} \par}
\caption{The $\ddot \nu - \dot \nu$ diagram with the simple variations 
model. The solid line is the amplitude $A_{\ddot\nu} \equiv \sigma_{\ddot\nu}$
of the frequency 
second derivative variations, the dashed line is the secular term $\ddot\nu_{ev}$,
and the dot-dashed lines are the envelopes of the oscillations 
$\ddot\nu_{ev}\pm \sigma_{\ddot\nu}$ with 1-$\sigma$ ranges (dotted lines). 
Each pulsar spends the majority of its lifetime at or very near the 
envelopes. The pulsars are labeled as on Fig. \ref{fig_diagram}}
\label{fig_model}
\end{figure}

From Eq. (\ref{eq2}) we may easily determine the relation between
$\ddot\nu_{ev}$ 
and $\dot\nu$ as 
\begin{equation}
\ddot\nu_{ev} = nC^{\frac1n}(-\dot\nu)^{2-\frac1n},
\end{equation}
which is shown in Fig. \ref{fig_model} as a thick dashed line. The same relation may be 
also estimated directly by using the asymmetry of the branches seen on Fig. 
\ref{fig_diagram} as 
\begin{equation}
{\ddot\nu(\dot\nu) = \frac12\left(\ddot\nu_{+} + \ddot\nu_{-}\right)}, 
\end{equation}
where $\ddot \nu_{\pm}$ are defined in Fig. \ref{fig_diagram}. Such estimation, 
while being very noisy, is positive in the 
$-10^{-11} \div -10^{-15}$ s$^{-2}$ range and agrees quantitatively with 
the previous one.

The amplitude of the $\ddot \nu$ oscillations, $A_{\ddot \nu}$, may be easily 
computed in a similar way, by using  $\ddot\nu_{+}$ and $\ddot\nu_{-}$, as 
$A_{\ddot \nu} = \frac12(\ddot \nu_{+} - \ddot\nu_{-})$.

The behaviour of pulsars according to the derived relations is shown in Fig. 
\ref{fig_model}. This simple variations model describes the observed branches, 
both positive and negative, rather well. We interpret the absence of negative branch 
objects with $\dot\nu<-10^{-11}$ s$^{-2}$, i.e. with $\tau_{ch}<10^4$, as a
prevalence of the second derivative's secular component over 
the varying one ($A_{\ddot\nu} < \ddot\nu_{ev}$) in this region. Older pulsars 
begin to change the sign of $\ddot\nu$ because of spin rate variations.

\section{Discussion and conclusions}
\label{disc}
In general, it is impossible to estimate the amplitudes of the frequency and 
its first derivative variations $A_{\nu}$ and $A_{\dot\nu}$ from the amplitude
of the second derivative only (the knowledge of its complete power density 
spectrum is needed). However, if the spectral density is relatively localized 
and some characteristic timescale $T$ of the variations exists, it is possible
to set some limits on it. A rough estimation is 
$A_{\nu}\sim A_{\dot\nu}T$, $A_{\dot\nu}\sim A_{\ddot\nu}T$ and 
$A_{\nu}\sim A_{\ddot\nu}T^2$. On a long timescale the variations can not 
lead to pulsar spin-up, and therefore, the variations of frequency first derivatives are much smaller than their secular values, and 
${A_{\dot\nu}\sim A_{\ddot\nu}T \ll \dot\nu}$, so $T\ll\dot\nu/A_{\ddot\nu}$.
So, for the $-10^{-12} < \dot\nu < -10^{-15}$ s$^{-2}$ range and corresponding
values of $A_{\ddot\nu}$ from $10^{-23}$ s$^{-3}$ to $10^{-26}$ s$^{-3}$, the 
characteristic timescale $T_{up} \sim 10^{11}$ s. Also, this characteristic 
timescale is obviously larger than the timespan of observations:
$50 < T < 3 \cdot 10^3$ years. Assuming the constancy of $T$ during the pulsar 
evolution and therefore the change of $A_{\nu}$ with time, we get
$A_{\nu} \sim 10^{-3} \div 10^{-7}$ Hz. For such a model the pulsar 
frequency varies with the characteristic time of several hundred years and 
the amplitude from $10^{-3}$ Hz for young objects to $10^{-7}$ Hz for older 
ones.

\begin{figure}[t]
{\centering \resizebox*{1\columnwidth}{!}{\includegraphics[angle=270]
{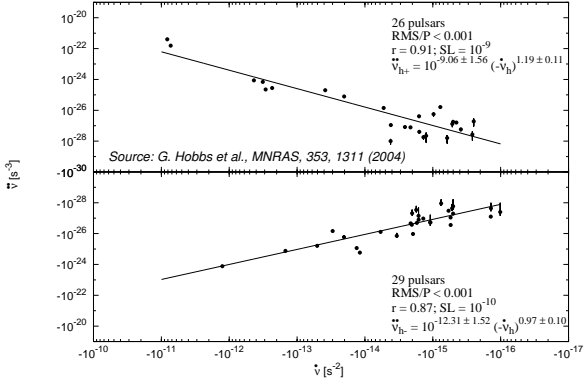}} \par}
\caption{The $\ddot \nu - \dot \nu$ diagram for low noise pulsars.
{This diagram shows a possible difference between long timescale variations,
discussed in the current work and the well-known ``timing-noise'', because
some of the pulsars (especially with low $|\dot \nu|$) still have anomalous 
braking indices.}}
\label{fig_lownoise}
\end{figure}

The characteristic timescale $T$ measured by the method above does not depend on
the derived value of $n \approx 5$. Therefore, even if the braking index is
significantly bigger than $5$, the values of $T$ and $A_{\nu}$ will not
be changed. 

The physical reasons of the discussed non-monotonous variations of the pulsar
spin-down rate may be similar to the ones of the timing noise on a short 
timescale. Several processes had been proposed for their explanation 
\citep{cor81} -- from the collective effects in the neutron star superfluid 
core to the electric current fluctuations in the pulsar magnetosphere. 
Whether these processes are able to produce long timescale variations is 
yet to be analysed. On a short timescale, the pulsars show different timing 
behaviour. But on the long timescale their behaviour seems to be alike.

The argument in favour of the similarity between the discussed variations and
the timing noise is the coincidence of the timing noise $\ddot\nu$ amplitude
extrapolated according to its power spectrum slope \citep{bay99} to the timescale of hundreds of years, with the $A_{\ddot\nu}$ derived from our analysis 
for the same $\dot\nu$, i.e. the same ages.

\begin{figure}[t]
{\centering \resizebox*{1\columnwidth}{!}{\includegraphics[angle=270]
{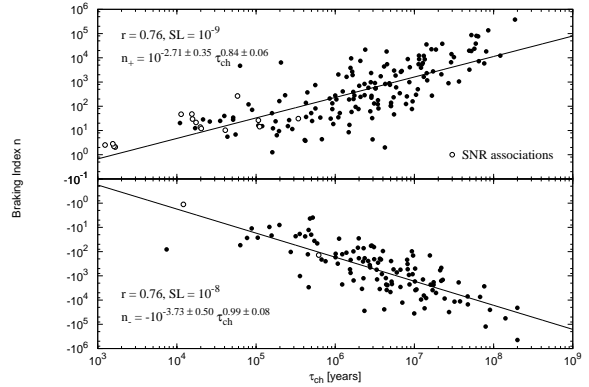}} \par}
\caption{The $n - \tau_{ch}$ diagram. The older pulsars have a larger spread
of braking index values. But some of them show a very low timing noise 
within the timespan of observation (see Fig. \ref{fig_lownoise}).}
\label{fig_ntau}
\end{figure}

At the same time, there are several low noise pulsars with large or negative
$\ddot\nu$ (see Figs. \ref{fig_lownoise} and \ref{fig_ntau}). For 55 pulsars
studied in \citep{hob04} the timing noise 
is nearly absent (RMS $< 1\cdot10^{-3} P$). Some of them have anomalous values of
$\ddot\nu$ , which are well consistent with the 
$|\ddot\nu|$ - $\dot \nu$ correlation \citep{cor85, arz94} and have a wide 
range of $\dot \nu$. This shows a possible difference between timing noise 
and the long timescale variations described above.

In any case, the principal point is that all the pulsars evolve with
long-term variations, and the timescale of such variations significantly exceeds 
several tens of years. That explains the anomalous values of the observed 
$\ddot\nu$ and braking indices and gives reasonable values of the underlying 
secular spin-down parameters. 



\begin{thebibliography}{3}

\bibitem[\protect\citeauthoryear{Arzoumanian et al.}{1994}]{arz94}
Arzoumanian, Z., Nice, D. J., Taylor, J. H. et al. 
Timing behavior of 96 radio pulsars. ApJ, 422, 671--680, 1994

\bibitem[\protect\citeauthoryear{Baykal et al.}{1999}]{bay99}
Baykal, A., Ali Alpar, M., Boynton, P. E. et al. 
The timing noise of PSR 0823+26, 1706-16, 1749-28, 2021+51 and the 
anomalous brakingindices. 
MNRAS, 306, 207--212, 1999

\bibitem[\protect\citeauthoryear{Beskin et al.}{1993}]{bes93}
Beskin, V., Gurevich, A. \& Istomin, Ya.
Physics of the Pulsar Magnetosphere, Cambridge: Cambridge University Press, 1993

\bibitem[\protect\citeauthoryear{Chukwude}{1993}]{chu03}
Chukwude, A.E. 
On the statistical implication of timing noise
for pulsar braking index.
A\&A, 406, 667--671, 2003

\bibitem[\protect\citeauthoryear{Cordes \& Chernoff}{1998}]{cor98}
Cordes, J. M. \& David F. Chernoff
Neutron Star Population Dynamics. II. Three-dimensional Space Velocities of
Young Pulsars
ApJ, 505, 315--338, 1998

\bibitem[\protect\citeauthoryear{Cordes \& Downs}{1985}]{cor85}
Cordes, J. M. \& Downs, G. S. 
JPL pulsar timing observations. III - Pulsar rotation fluctuations. 
ApJS, 59, 343--382, 1985

\bibitem[\protect\citeauthoryear{Cordes \& Greenstein}{1981}]{cor81}
Cordes, J. M. \& Greenstein, G. 
Pulsar timing. IV - Physical models for timing noise processes. 
ApJ, 245, 1060--1079, 1981


\bibitem[\protect\citeauthoryear{D'Alessandro et al.}{1993}]{d'a93}
D'Alessandro, F., McCulloch, P. M., King, E. A. et al. 
Timing Observations of Southern Pulsars - 1987 TO 1991. 
MNRAS, 261, 883--894, 1993

\bibitem[\protect\citeauthoryear{Davis \& Goldstein}{1970}]{dav70}
Davis, L., Goldstein, M.
Magnetic-Dipole Alignment in Pulsars.
ApJ, 159, L81--L86, 1970

\bibitem[\protect\citeauthoryear{Demia\'{n}sky \& Proszy\'{n}ski}{1979}]{dem79}
Demia\'{n}sky, M.,  Proszy\'{n}ski, M.
Does PSR0329+54 have companions?
Nature, 282, 383--385, 1979

\bibitem[\protect\citeauthoryear{Goldreich \& Julian}{1969}]{gj69}
Goldreich, P., Julian, W. H.
Pulsar Electrodynamics.
ApJ, 157, 869--880, 1969

\bibitem[\protect\citeauthoryear{Gullahorn \& Rankin}{1977}]{gul77}
Gullahorn, G. E., Rankin, J. M.
Second Derivatives of Pulsar Rotation Frequencies.
Bull. of the Am. Astro. Soc., 9, 562, 1977


\bibitem[\protect\citeauthoryear{Hobbs et al.}{2004}]{hob04}
Hobbs, G., Lyne, A. G., Kramer, M. et al. 
Long-term timing observations of 374 pulsars.
MNRAS, 353, 1311--1344, 2004

\bibitem[\protect\citeauthoryear{Krolik}{1991}]{kro91}
Krolik, J.H.,
Multipolar magnetic fields in neutron stars.
ApJL, 373, 69--72, 1991

\bibitem[\protect\citeauthoryear{Lyne}{1999}]{lyn99}
Lyne, A. in: Arzoumanian Z., van der Hooft F. \&
van den Heuvel E. P. J. (Eds.) Pulsar Timing, General Relativity and 
the Internal Structure of Neutron Stars, 
Amsterdam, p.141, 1999

\bibitem[\protect\citeauthoryear{Manchester et al.}{2005}]{man05}
Manchester, R. N., Hobbs, G. B., Teoh, A. et al. 
The Australia Telescope National Facility Pulsar Catalogue.
AJ, 129, 1993--2006, 2005

\bibitem[\protect\citeauthoryear{Manchester \& Taylor}{1977}]{man77}
Manchester, R. N. \& Taylor, J. H. 
Pulsars, San Francisco: Freeman, 1977

\bibitem[\protect\citeauthoryear{Ruderman}{2006}]{rud06}
Ruderman, M.
Pulsar Spin, Magnetic Fields, and Glitches.
astro-ph/0610375

\bibitem[\protect\citeauthoryear{Shemar \& Lyne}{1996}]{she96}
Shemar, A. L. \& Lyne, A. G. 
Observations of pulsar glitches. 
MNRAS, 282, 677--690, 1996

\bibitem[\protect\citeauthoryear{Stairs et al.}{2000}]{sta00}
Stairs, I. H., Lyne, A. G. \& Shemar, A. L. 
Evidence for free precession in a pulsar. 
Nature, 406, 484--486, 2000

\bibitem[\protect\citeauthoryear{Urama et al.}{2006}]{ura06}
Urama, J.O., Link, B., \& Weisberg, J. M. 
A strong $\ddot \nu - \dot \nu$ correlation in radio pulsars with
implications for torque variations.
MNRAS, 370, L76--L79, 2006

\bibitem[\protect\citeauthoryear{Zane \& Turolla}{2006}]{zane06}
Zane, S., Turolla, R.
Unveiling the thermal and magnetic map of neutron star surfaces though their
X-ray emission: method and light-curve analysis.
MNRAS, 366, 727--738, 2006


\end{thebibliography}
\end{document}